
\input harvmac
\def\trr{\tr R^2}
\def\trf#1{\tr F_{#1}^2 }


\def\mpl#1#2#3{Mod. Phys. Lett. {\bf A#1} (#2) #3}

\lref\sminst{E. Witten, ``{\it Small Instantons in String Theory,}''
IASSNS-HEP-95-87, hep-th/9511030.}

\lref\dbrane{J. Polchinski,
``{\it Dirichlet-Branes and Ramond-Ramond Charges,}'' hep-th/9510017;
J. Dai, R.G. Leigh and J. Polchinski,
``{\it New Connections between String Theories,}'' \mpl{4}{1989}{2073};
R.G. Leigh, ``{\it Dirac-Born-Infeld Action from Dirichlet $\sigma$-model,}''
\mpl{4}{1989}{2767}.}

\lref\rutgrp{T. Banks, M. Berkooz, M.R. Douglas, R.G. Leigh, N. Seiberg, et al,
work in progress.}

\lref\gsw{M.B. Green, J.H. Schwarz and E. Witten,
``{\it Superstring Theory,}''  Cambridge University Press, 1987.}

\lref\erler{J. Erler, ``{\it Anomaly Cancellation in Six Dimensions,}''
J. Math. Phys. {\bf 35} (1994) 1819.}

\lref\johns{J.H. Schwarz, ``{\it Anomaly-Free Supersymmetric Models in
Six Dimensions,}'' CALT-68-2030, hep-th/9512053.}

%
\def\boxit#1{\vbox{\hrule\hbox{\vrule\kern3pt
\vbox{\kern3pt#1\kern3pt}\kern3pt\vrule}\hrule}}
\newdimen\str
\def\fboxit#1#2{\vbox{\hrule height #1 \hbox{\vrule width #1
\kern3pt \vbox{\kern3pt#2\kern3pt}\kern3pt \vrule width #1 }
\hrule height #1 }}
\def\yboxit#1#2{\vbox{\hrule height #1 \hbox{\vrule width #1
\vbox{#2}\vrule width #1 }\hrule height #1 }}
\def\fillbox#1{\hbox to #1{\vbox to #1{\vfil}\hfil}}
\def\dotbox#1{\hbox to #1{\vbox to 8pt{\vfil}\hfil $\cdots$ \hfil}}

\def\yboxs{\yboxit{0.4pt}{\fillbox{5pt}}\hskip-0.4pt}

\def\tableaux#1{\vcenter{\offinterlineskip \halign{&\tabskip 0pt##\cr
#1}}\ }
\def\cropen#1{\crcr\noalign{\vskip #1}}
\def\cry{\cropen{-0.4pt}}

\Title{RU-95-94,  hep-th/9512191}
{\vbox{\centerline{Anomalies, D-flatness and Small Instantons}}}
\bigskip
\bigskip
\centerline{R.G. Leigh}
\bigskip
\centerline{\it Department of Physics and Astronomy}
\baselineskip=14pt
\centerline{\it Rutgers University}
\baselineskip=14pt
\centerline{\it Piscataway, NJ 08855-0849}
\baselineskip=14pt
\centerline{\tt leigh@physics.rutgers.edu}
\bigskip\smallskip
\vglue 0.8cm
\centerline{\rm ABSTRACT}
\bigskip
\baselineskip 18pt
\noindent

Recently, Witten has proposed a mechanism for symmetry enhancement in
$SO(32)$ heterotic string theory, where the singularity obtained by
shrinking an instanton to zero size is resolved by the appearance of
an $Sp(1)$ gauge symmetry. In this short letter, we consider spacetime
constraints from anomaly cancellation in six dimensions and D-flatness
and demonstrate a subtlety which arises in the moduli space when
many instantons are shrunk to zero size.

\Date{December 1995}

In a recent paper, Witten\sminst\ has described a new mechanism for
enlarging the rank of the gauge group of field theories arising from
$SO(32)$ heterotic strings. One considers a six-dimensional
compactification on a $K3$ surface, which can be taken to be at a generic
point in moduli space. From the index theorem, we know that we must
choose a nontrivial gauge bundle $V$ with instanton number $c_2(V)=24$.
There are many ways to do this, and one can describe the situation in
terms of an $SO(N)$ gauge configuration, for some $4\leq N\leq 24$. More
precisely, the structure group of an instanton is $SU(2)$ and there
are 24 such instantons embedded in the $SO(32)$ group
in some way. The details of this embedding determine the unbroken gauge
group, or at least the part described by conformal field theory.

In Ref. \sminst, Witten considered the singularity obtained when
the instanton scale size approaches zero. It was argued that this
singularity may be resolved in terms of an enhanced $Sp(1)$ gauge
symmetry. The full gauge group then consists of this $Sp(1)$ factor
times some subgroup of $SO(32)$, determined by the nature of the
remaining gauge bundle. This may be done to any number of the
instantons and in the special case where all 24 instantons have
been shrunk to zero size, the gauge group is $SO(32)\times Sp(1)^{24}$.
This may be enhanced to $SO(32)\times Sp(24)$ by bringing the
positions of the point instantons together.
As Witten showed, this has a satisfying Type I dual description in terms
of D-branes.\dbrane\ For each point instanton, there is a 5-brane which
carries $Sp(1)$ Chan-Paton factors.\foot{The world-brane field theory contains
$Sp(1)$ super Yang Mills.} As $k$ of the 5-branes come together,
there are additional light Dirichlet open string states which fill out
a vector multiplet of $Sp(k)$.

Having compactified on $K3$, the low-energy field theory possesses a
6-dimensional $N=1$ sypersymmetry.\foot{Corresponding to $N=2$ in
$D=4$.} It is well known that anomaly cancellation in 6 dimensions is
quite restrictive. In this letter, we wish to enlarge on the discussion
of Ref. \sminst\ paying particular attention to anomaly cancellation
and D-flatness.
We will reconstruct the spectrum of the theory of
maximal symmetry and demonstrate its (Green-Schwarz) anomaly
cancellation. Next, we study the moduli space by solving the D-flatness
constraints. These constraints turn out to be somewhat non-trivial and
indicate
which field theories may be accessed by Higgsing. Of course, Higgsing
(along a flat direction!) must not spoil the anomaly cancellation and we
demonstrate that this is so. The resulting theories may be given an
interpretation in terms of D-branes: these theories may be
understood\rutgrp\ as consistent compactifications of open strings on
$K3$ orbifolds.

\newsec{Anomaly Cancellation}

The theory of maximal symmetry given by Witten has gauge group
$SO(32)\times Sp(24)$. The hypermultiplets arrange themselves in the
representations
\eqn\maxsym{\eqalign{
ND:&\;\;\;{1\over 2}\times ({\bf 32,48})\cr
DD:&\;\;\; \;\;\; ({\bf 1,}\; \tableaux{\yboxs\cry\yboxs\cry})\cr
K3:&\;\;\; 20\times ({\bf 1,1}).\cr
}}
The labels such as `ND' refer to the labeling of states\foot{We are
counting full hypermultiplets. The factor
$1/2$ in the first line of eq. \maxsym\ refers to the fact that the
irreducible supersymmetry representation of a field in a pseudoreal
representation of the gauge group contains one independent bosonic
field (on-shell).} in the D-brane
representation. The 20 singlet states may be understood as the 4
hypermultiplets of the internal $g_{ij}, B_{ij}$ plus the 16 blowing-up
modes. The total number of hypermultiplets ($n_H=768+1128+20=1916$) is 244
more than the number of vector multiplets, and thus the $\tr R^4$
anomaly cancels.{\gsw,\erler}
In addition, the $\tr F^4$ anomaly terms cancel. The remaining
mixed gravitational-gauge anomalies will not cancel, but can be
eliminated by the Green-Schwarz mechanism provided they factorize in a
suitable way. In the present case, we find (see also Ref. \johns)
\eqn\maxsymmanom{
I \propto (\trr +2\trf{32}-2\trf{24})\times (\trr-\trf{32})
}
and so indeed the model is anomaly-free.

By turning on the antisymmetric $DD$ field, we may break the $Sp(24)$
to a product of symplectic groups with total rank 24. This Higgsing
is always D-flat, and we will for simplicity now consider the
$SO(32)\times Sp(1)^{24}$ theory with hypermultiplets
\eqn\thrtwo{\eqalign{
{1\over 2}\times &({\bf 32,2,1}^{23})\;\; +{\rm permutations}\cr
44\times &({\bf 1,1}^{24} )\cr
}}
where the 44 singlets come from the 20 original singlets plus 24
Higgs modes which are uneaten. One may easily check that the
anomalies cancel, with the mixed anomalies now factorizing as
\eqn\thrtwomanom{
I \propto \left(\trr +2\trf{32}-2\sum_i\trf{i}\right)\times
\left(\trr-\trf{32}\right).
}

In the remainder of the paper, we will consider theories obtainable
through consistent Higgsing of this model.

\newsec{D-flatness}

Several general aspects of the flat directions were discussed in
Ref. \sminst. The $D=6$, $N=1$ supersymmetry algebra contains an
$SU(2)_R$ symmetry and the D-terms transform in the triplet representation:
\eqn\gendterm{\eqalign{
D^a_{ij}&= \sum_k T^a_{AB} \phi^{A\alpha_k}_i\phi^{B\beta_k}_j
\epsilon_{\alpha\beta}\cr
D^{\tilde a_k}_{ij} &= T^{\tilde a_k}_{\alpha_k\beta_k}
\phi^{A\alpha_k}_i\phi^{A\beta_k}_j
}}
for $k=1,\ldots, 24$. Here, $i,j$ are $SU(2)_R$ indices, $a;A,B$ are
$SO(32)$ indices, $\tilde a_k; \alpha_k,\beta_k$ are indices of the
$k$th $Sp(1)$ factor and $\phi$ are the non-singlet fields of eq. \thrtwo.
In the case at hand $\phi$ will also satisfy a reality condition
\eqn\reality{
\overline{\phi}^{A,i}_{\alpha_k}=\epsilon_{\alpha_k\beta_k}\epsilon^{ij}
\phi^{A\beta_k}_j.}
The multiplicity of D-term conditions may be simply understood in terms
of the D- and F-flatness conditions of the corresponding dimensionally
reduced $D=4$, $N=2$ theory.

We consider the case where $SO(28)$ is unbroken; the more general case
will be considered shortly. The general picture advocated in Ref.
\sminst\ is that the shrinking of instanton scale sizes should be
understood in terms of the appearance of extra $Sp(1)$ factors. The
converse is also true, but with some important caveats.
In particular, if we start with the $SO(32)\times Sp(1)^{24}$ theory
given above, we cannot unshrink a {\it single} instanton: this is not a
flat direction of the theory. In fact, the minimal operation is an
unshrinking of {\it four} instantons, as we will now show.

This may be demonstrated by explicitly examining the D-terms when
just one of the $Sp(1)$'s is Higgsed. One finds that there are no
flat directions satisfying all of the constraints \gendterm.  Indeed,
focus on a single such field $\phi$, which is charged, say, under the
first $Sp(1)$
factor. Using just the reality condition and the $Sp(1)$ D-flatness condition,
one finds nine real conditions on 14 real parameters (recall that we are
assuming that $SO(28)$ is unbroken, and we have used our freedom of
$Sp(1)$ transformations to eliminate one complex parameter. For later
use (where more than one vev will be considered), it will be convenient
to not perform $SO(4)\subset SO(32)$ rotations. Thus the solutions to
those equations can be parameterized by 5 real parameters (some of
which are gauge redundant in the present case). However, we must also
satisfy the
$SO(4)\subset SO(32)$ D-flatness conditions. It can easily be seen
that these constitute $6\times 3=18$
real conditions, and we do not expect a solution; this can be born out
by direct computation.

If one turns on $n$ such fields, the $Sp(1)$
D-flat solutions are parameterized by $5n$ real parameters. Restricting
to those solutions which are also $SO(4)$ D-flat leaves $5n-18$
parameters. Generically then, one must turn on at least $n\geq 4$ such
fields; the unbroken symmetry in such a vacuum is $SO(28)\times SU(2)
\times Sp(1)^{24-n}$. Four fields transforming as the ${\bf 28}$
of $SO(28)$ are eaten, as well as some $SO(28)$ singlets. We are
left with the following spectrum:
\eqn\spectrum{\eqalign{
{1\over 2}\times &({\bf 28,1;2,1}^{23-n})\;\; +{\rm permutations}\cr
{(n-4)\over 2}\times &({\bf 28,2;1}^{24-n})\cr
&({\bf 1,2;2,1}^{23-n})\;\; +{\rm permutations}\cr
{(n-3+44)}\times &({\bf 1,1;1}^{24-n})\cr
}}
The total number of hypermultiplets is
$n_H=56(24-n)+28(n-4)+4(24-n)+(n+41)=697-3n$, and there are
$n_V=378+3+3(24-n)=453-3n$ vectors. Thus, the $\tr R^4$ anomaly has
cancelled as it must. Incidentally, the requirement
of $n\geq 4$ can also be
seen through the $\tr F_{28}^4$ anomaly: one must have precisely
20 fields transforming as the ${\bf 28}$ of $SO(28)$; not enough
such fields can be eaten until $n\geq 4$. The mixed anomalies
can be shown to factorize as
\eqn\specnmanom{
I\propto \left(\trr+2\trf{28}-2(n-2)\trf{2}-2\sum_{i=1}^{24-n}\trf{i}\right)
\times\left(\trr-\trf{28}-2\trf{2}\right)}
and thus can be cancelled by the Green-Schwarz mechanism.
The spectrum for $n>4$ can also be obtained from the $n=4$ case
by turning on some of the $({\bf 1,2;2,1}^{19})$ fields present in
that case.

\newsec{$SO(N)$ bundles}

We may also consider $SO(N)$ bundles for $N>4$. For given $N$ and $n$,
there are
two branches, one where the gauge symmetry is $SO(32-N)\times Sp(1)^{24-n}$,
the other where there is an additional $SU(2)$. We consider the first case.
The D-flatness conditions lead us to turn on $n\geq N$ vevs, and the following
spectrum may be shown to be anomaly-free:
\eqn\nspec{\eqalign{
{1\over 2}\times &({\bf 32-N;2,1}^{23-n})\;\; +{\rm permutations}\cr
{(n-N)}\times &({\bf 32-N;1}^{24-n})\cr
{N\over 2}\times &({\bf 1;2,1}^{23-n})\;\; +{\rm permutations}\cr
{(44-n+x_n)}\times &({\bf 1;1}^{24-n})\cr
}}
where $x_n=n(N-2)-{1\over 2}N(N-1)$. This number is exactly what is required
to interpret this Higgsing in terms of the unshrinking of
small instantons: $x_n$ is precisely the dimension of the moduli space
of $n$ instantons, that is, the number of deformations of an $SO(N)$
gauge bundle of instanton number $n$. We count as follows: each
instanton comes with a scale size, 4 position coordinates and 3 $SU(2)$
orientations. The possible embeddings
of $SU(2)$ in $SO(N)$ are parameterized by the coset
$SO(N)/(SO(N-4)\times SO(4)$
of (real) dimension $4(N-4)$. This corresponds to a total of
$(N-2)$ hypermultiplets for each instanton. Subtracting off the overall
$SO(N)$ gauge rotations, we arrive at  $n(N-2)-{1\over2}N(N-1)$ as the
dimension of the n-instanton moduli space. The counting of singlet fields
is also nicely interpreted in terms of D-branes: when the vev of a given
field is zero, there is a 5-brane, with a single hypermultiplet describing
its position on the compact space. All other fields are in non-trivial
representations of the gauge group. When the corresponding vev is turned on
(i.e. an instanton unshrunk), the 5-brane disappears (accounting for the
$-n$ in the number of singlets), and $x_n$ moduli describe
the resulting finite-size instanton.

There is an additional branch of the moduli space where an enhanced $SU(2)$
gauge symmetry arises.  In this case, we find the anomaly-free spectrum
in representations of $SO(32-N)\times SU(2)\times Sp(1)^{24-n}$:
\eqn\nspecsutwo{\eqalign{
{1\over 2}\times &({\bf 32-N,1;2,1}^{23-n})\;\; +{\rm permutations}\cr
{(n-N)\over 2}\times &({\bf 32-N,2;1}^{24-n})\cr
&({\bf 1,2;2,1}^{23-n})\;\; +{\rm permutations}\cr
{(N-4)\over 2}\times &({\bf 1,1;2,1}^{23-n})\;\; +{\rm permutations}\cr
z\times &({\bf 1,2;1}^{24-n})\;\; +{\rm permutations}\cr
{(44-n+x_n-2z)}\times &({\bf 1,1;1}^{24-n})\cr
}}
where $z=(N-4)(n-N+4)/2$. Some of the singlets of
eq. \nspec\ now transform as doublets of the $SU(2)$.

\newsec{Conclusions}

The result of Section 2 is rather surprising given the 5-brane
interpretation of
Ref. \sminst. One may have suspected that the 5-branes can be removed
one by one; however, as we have seen, there are important spacetime
constraints on this procedure. In any case, we have clarified the
picture advocated by Witten, that the scaling of instanton size
corresponds to (un)Higgsing. Indeed,
this must be so, as this is the only possible
non-trivial dynamics at low energies in
six dimensions. Further properties of these and related models
will be explored in Ref. \rutgrp.


I wish to thank S. Chaudhuri for many useful conversations and inspiration,
and J.H. Schwarz for conversations
regarding some of the results presented early in section 1. I also
thank many members of the Rutgers string group for ongoing
collaboration and E. Witten for comments.
Research supported in part by the U.S. Department of Energy, contract
DE-FG05-90ER40599.

\listrefs \end